\documentstyle[prl,aps,multicol,amsmath,graphicx,psfrag,color]{revtex}

\begin{document}

\newcommand{\fig}[2]{\includegraphics[width=#1]{#2}}

\draft
\title{Momentum-resolved tunneling between Luttinger liquids} 
\author{D. Carpentier$^1$, C. Pe\c ca$^2$, and L.~Balents$^2$}
\address{$^1$~Institute for Theoretical Physics, University of
  California, Santa Barbara, CA 93106 \\ ${}^2$~Physics Department,
  University of California, Santa Barbara, CA 93106}

\date{\today}
\maketitle

\begin{abstract}
  We study tunneling between two nearby cleaved edge quantum wires in
  a perpendicular magnetic field.  Due to Coulomb forces between
  electrons, the wires form a strongly-interacting pair of Luttinger
  liquids.  We calculate the low-temperature differential tunneling
  conductance, in which singular features map out the dispersion
  relations of the {\sl fractionalized} quasiparticles of the system.
  The velocities of several such {\sl spin-charge separated}
  excitations can be explicitly observed.  Moreover, the proposed
  measurement directly demonstrates the splintering of the tunneling
  electrons into a multi-particle continuum of these quasiparticles,
  carrying separately charge from spin.  A variety of corrections to
  the simple Luttinger model are also discussed.
\end{abstract}

\begin{multicols}{2}
  
  The only universally accepted example of a non-Fermi liquid metallic
  state is the one dimensional (1D) {\sl Luttinger liquid}
  (LL)\cite{Haldane81}.  Remarkably, the quasiparticle excitations of
  a LL are {\sl fractionalized}, comprising a diverse set carrying
  spin separately from charge, and charge in fractions of the electron
  charge $e$.  LL behavior {\sl has} been observed experimentally in
  carbon nanotubes\cite{Bockrath99}, through strongly energy dependent
  {\sl local} tunneling, and more recently in GaAs quantum wires
  through power-law resonant tunneling lineshapes\cite{Auslaender}.
  Evidence of charge fractionalization has also been seen in shot
  noise experiments using fractional quantum Hall edge
  states\cite{noise1}, which are somewhat special {\sl chiral}
  Luttinger liquids\cite{edge}.  Despite these successes, no direct
  experimental evidence of fractionalization has ever been obtained in
  a non-chiral 1D system.  In this letter, we show that measurements
  of the non-linear tunneling conductance between parallel Luttinger
  liquids in a transverse magnetic field provide a direct {\sl
    spectroscopic} probe of fractionalization.  The results, as
  described below, give very similar information to an {\sl ideal}
  photoemission experiment.  Indeed, some indications of spin-charge
  separation were seen in photoemission spectroscopy of the quasi-1D
  cuprate SrCuO$_2$\cite{Kim96}.  Tunneling spectroscopy has, however,
  the advantages of being possible on a single, {\sl isolated} 1D
  system and with potentially much higher resolution than
  photoemission.

Controlled tunneling experiments between two parallel wires have been 
recently conducted using  cleaved-edge overgrowth by O.M. Auslaender 
{\sl et al}\cite{Yacoby}.
The experimental geometry we consider 
is indicated schematically in
Fig.~1.  The two ``wires'' are in fact confined surface states, and
electrical contact is made only to the upper wire via a
Two-Dimensional Electron Gas (2DEG).  With $L' \gg L$, nearly the full
electrochemical potential drop occurs between the shorter (left) segment of the
upper wire and the lower wire.  Because of the uniformity of the
barrier, momentum along the wire is conserved during
tunneling\cite{Kang00}.
\begin{figure}[hbt]
\psfrag{L1}{\bf L}
\psfrag{L2}{\bf L'}
\centerline{\fig{9cm}{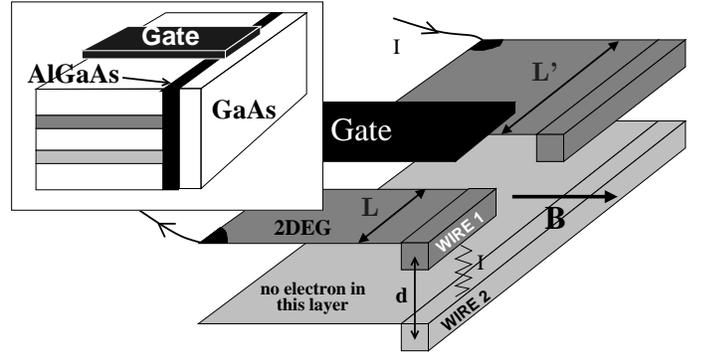}}
\caption{Schematic experimental geometry (from Ref.~7)
}
\label{fig:experience}
\end{figure}

We assume that
the barrier is sufficiently high as to establish a quasi-equilibrium
state on either side of the barrier, treating tunneling across the
barrier perturbatively.  
A non-interacting model Hamiltonian for the system neglecting tunneling is then
\begin{equation}
  H_0 = \sum_{a\alpha} \int\!\! dx\; c_{a\alpha}^\dagger \left[ - \frac{\partial_x^2}{2m} - U_a - \mu_a\right]c_{a\alpha}^{\vphantom\dagger} ,
  \label{eq:H0}
\end{equation}
where $a=1,2$ labels the upper/lower wire,
$\alpha=\uparrow,\downarrow$ labels the electron spin, and $U_a$ and
$\mu_a$ are the electrostatic and chemical potential of the $a^{th}$
wire, respectively.  We choose by convention to take $\mu_a=0$ in
equilibrium, so that $k_{{\rm \scriptscriptstyle F}a} = \sqrt{2m U_a}$
is the Fermi momentum in wire $a$.  Neglecting the energy dependence
of the tunneling amplitude $w$, the Zeeman shift (see below), and a small
energy shift due to orbital magnetic effects {\sl within} each wire,
the tunneling Hamiltonian in the presence of a magnetic field
$B\hat{z}$ (in the gauge $A_y = Bx$) is
\begin{equation}
  H_{\rm tun} = - w\sum_\alpha \int\!\! dx\; \left[
    c_{1\alpha}^\dagger c_{2\alpha}^{\vphantom\dagger}e^{iQx} +
    c_{2\alpha}^\dagger c_{1\alpha}^{\vphantom\dagger}e^{-iQx} \right].
\end{equation}

\begin{figure}[thb]
\psfrag{m2=0}{\hspace{-.2 cm}$\mu_{2}=0$}
\psfrag{m1=}{\hspace{-0.2 cm}$\mu_{1}=$}
\psfrag{V}{$V$}
\psfrag{Q}{$Q$}
\centerline{\fig{5cm}{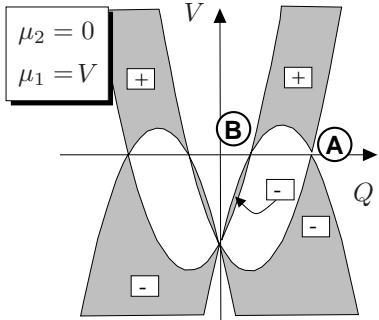}}
\caption{Tunneling current in the $V-Q$ plane for the non-interacting 
  model.  Regions of non-zero current are shaded, and $dI/dV$ has
  delta-function singularities along the boundaries between shaded and 
  unshaded regions.  }
\label{fig:nointpd}
\end{figure}

Here the magnetic wavevector $Q=2\pi B d/\phi_0$, $d$ is the
center-to-center distance of the wires, and $\phi_0 = hc/e$ is the
flux quantum.  
The one-dimensional (1D) tunneling current density, $J = i e w
(c_{1\alpha}^\dagger c_{2\alpha}^{\vphantom\dagger}e^{iQx} -
c_{2\alpha}^\dagger c_{1\alpha}^{\vphantom\dagger}e^{-iQx} )$ can be
calculated directly from Fermi's
golden rule.  The result, whose gross features appear experimentally
in Ref.~\onlinecite{Yacoby},  is shown in Fig.~\ref{fig:nointpd}, taking
$\mu_1=V$, $\mu_2=0$.  This diagram can be understood physically by
considering all processes by which an electron can be transferred
between the two wires, moving from an occupied to an unoccupied state.
Geometrically, Fig.~2 is obtained by drawing the locus of values of
$V,Q$ for which one of the four Fermi points lies on the other
parabola when shifted vertically and horizontally by $V$ and $Q$,
respectively.  

{\sl Zero bias} features occur when (A) $Q=\pm(k_{{\rm\scriptscriptstyle
    F}1} + k_{{\rm\scriptscriptstyle F}2})$, (B) $Q=\pm
(k_{{\rm\scriptscriptstyle F}1} - k_{{\rm\scriptscriptstyle F}2})$.
It is in the {\sl low-energy} region of the $V-Q$ plane near these
four features that {\sl universal} features arise in the {\sl
  interacting} system, to which we now turn.
Focusing on the the low-bias regime, we first decompose the electron
fields into right and left movers, $c_{a\alpha}(x) =  c_{Ra\alpha}(x)
e^{i k_{{\rm\scriptscriptstyle F}a} x} + c_{La\alpha}(x)
e^{-i k_{{\rm\scriptscriptstyle F}a} x}$, 
which are described by the
Luttinger Hamiltonian (taking $\mu_a=0$ for simplicity)
\begin{equation}
  H_0 = - i\sum_a \int\!\! dx\; v_{{\rm\scriptscriptstyle F}a} \left[
    c_{Ra\alpha}^\dagger\partial_x c_{Ra\alpha}^{\vphantom\dagger} -
    c_{La\alpha}^\dagger\partial_x
    c_{La\alpha}^{\vphantom\dagger}\right],
  \label{eq:luttinger}
\end{equation} 
where $v_{{\rm\scriptscriptstyle F}a} = k_{{\rm\scriptscriptstyle
F}a}/m$.  In general, these right- and left- moving Fermions
undergo a diverse set of scattering processes mediated by the
Coulomb interaction (screened by the 2DEG).  A systematic study of
these terms\cite{inprep}\ reveals an important simplification
due to the experimental fact that the width $W$ 
of
a typical cleaved edge quantum wire is large compared to the Fermi
wavelength\cite{Yacoby} 
($k_{{\rm\scriptscriptstyle F}a}W \gg 1$).  We also expect the 2D
screening length $\lambda_{s} \gtrsim W$.  These properties imply
a strong suppression of two-electron backscattering 
processes.  We therefore adopt 
a {\sl forward scattering model}
retaining only the strongest (unsuppressed) interactions, 
\begin{equation}
  H_{\rm int} = \frac{1}{2}\sum_{ab} \int\!\! dx \; n_a (x)
  V_{ab} \,n_b(x), \label{ints}
\end{equation}
with $H=H_0 + H_{\rm tun} + H_{\rm int}$, and $n_a = \sum_\alpha
c_{Ra\alpha}^\dagger c_{Ra\alpha}^{\vphantom\dagger} +
c_{La\alpha}^\dagger c_{La\alpha}^{\vphantom\dagger}$.
Eq.~\ref{ints}\ neglects momentum dependence of the forward-scattering 
interactions, and is hence valid for 
$eV \lesssim \hbar v_{{\rm\scriptscriptstyle F}a}/\lambda_{s}$. 
The interactions
$V_{ab}$ can be roughly estimated as
$V_{11}\simeq V_{22}\simeq  (2e^{2}/\epsilon)\ln (2 \lambda_{s}/W)$, 
$V_{12}\simeq  (2e^{2}/\epsilon)\ln (\lambda_{s}/d)$
 where $e$ is the electron charge, $\epsilon$ the wires dielectric
constant.  Experimentally, $d-W \ll W$\cite{Yacoby}, so the
inter-wire interaction 
$V_{12}$ is not 
negligible. Thus the tunneling conductance does not in fact probe the 
spectral properties of two {\sl decoupled} 1DEGs.  Nevertheless, all
properties of the interacting Hamiltonian $H_0+H_{\rm int}$ can be
calculated {\sl exactly} by bosonization (LL theory).

$ $From now on we will focus on the current at zero temperature around
the point A of Fig.~\ref{fig:nointpd}. An analogous description can be
obtained around each of the low bias points of fig.~\ref{fig:nointpd}
\cite{inprep}.  To lowest order in perturbation in $w$, we can write 
the current density $J$
as the sum $J=2e|w|^{2}(J_{+}-J_{-})$, where $J_{+},J_{-}$ are
positive functions which satisfy (time-reversal symmetry)
$J_{+}(Q,V)=J_{-}(-Q,-V)$, and $J_{+}$ ($J_{-}$) is nonzero only for
$V>0$ ($V<0$).  Around point A, $q=Q-(k_{F_{1}}+k_{F_{2}})$ is small,
and 
\[
J_{+}
= \Re \int_{-\infty}^{+\infty}\!\! dx 
\int_0^{\infty}\!\!\! dt~ 
e^{i((V+i\delta)t+qx)}
C_{E}^{R\to L}(x,\tau\to\epsilon+it) \label{Jpform}
\]
 where $C_{E}^{\scriptsize R\to L}$ is the Euclidean correlation
function
$C_{E}^{R\to L}(x,\tau)=
\left<
c^{\dagger}_{R1}c^{\vphantom\dagger}_{L2}(x,\tau)
c^{\dagger}_{L2}c^{\vphantom\dagger}_{R1}(0,0) \right>$. 
LL theory then gives
\begin{multline}\label{def-CELR}
C_{E}^{R\to L}(x,\tau)=
\frac{a_0^{2\theta_{1}+2\theta_{2}-1}}{(2\pi)^{2}} 
\prod_{a=1,2} (v_{sa} \tau- \epsilon_{a}ix)^{-1/2} \\
(v_{ca} \tau -\epsilon_{a} ix)^{-\eta-\theta_{a}}
 (v_{ca} \tau+\epsilon_{a}ix)^{+\eta-\theta_{a}}
\end{multline}
where $\epsilon_{a}=(-1)^{a+1}$, and $a_0$ (of order $W$) is a small
distance cut-off for the LL description.  Fractionalization is evident 
formally in Eq.~\ref{def-CELR}\ through the multiple branch points
characterized by distinct charge and spin velocities, $v_{c{\rm\scriptscriptstyle 1/2}}$ and
$v_{s{\rm\scriptscriptstyle 1/2}}$. 
For two independent wires ($V_{12}=0$), it is well-known that
$\eta=1/4$ and $\theta_1,\theta_2 >1/4$.  In the present case, the
strong interactions between the wires make $\eta$  interaction-dependent, and 
$\theta_{1},\theta_{2}$ can take values smaller than $1/4$. Indeed in
the case of two coupled identical wires 
($v_{{\rm\scriptscriptstyle F}1}=
  v_{{\rm\scriptscriptstyle F}2}=
  v_{{\rm\scriptscriptstyle F}},V_{11}=V_{22}$), these exponents take the
values 
$\theta_{1}=1/(4\sqrt{1+2(V_{11}+V_{12})/(\pi v_{F})})$,  
$\theta_{2}=\sqrt{1+2(V_{11}-V_{12})/(\pi v_{F})}/4$, 
$\eta=0$. General but non-illuminating formula are postponed to 
\cite{inprep}.

Letting $x \rightarrow t u$ in Eq.~\ref{def-CELR}\ and
integrating over $t$ gives
\begin{equation}\label{I+integral}
J_{+}(Q,V)
\propto \Gamma(1-2\theta_{1}-2\theta_{2})\times 
\Re \int_{-\infty}^{+\infty} du ~ h(u),
\end{equation}
where the complex function $h(u)$ is defined by 
\[
h(u)=[\delta -i(V+qu)]^{2\theta_{1}+2\theta_{2}-1}
     C_{E}^{R\to L}(u,i+\epsilon).
\]

\begin{figure}[t]
\psfrag{vs2}{$v_{s{\rm\scriptscriptstyle 1}}$}
\psfrag{vc1}{$v_{c{\rm\scriptscriptstyle 1}}$}
\psfrag{vc2}{$v_{c{\rm\scriptscriptstyle 2}}$}
\psfrag{-vs1}{$-v_{c{\rm\scriptscriptstyle 1}}$}
\psfrag{-vc1}{$-v_{s{\rm\scriptscriptstyle 2}}$}
\psfrag{-vc2}{$-v_{c{\rm\scriptscriptstyle 2}}$}
\psfrag{-V/q}{$-V/q$}
\centerline{\fig{9cm}{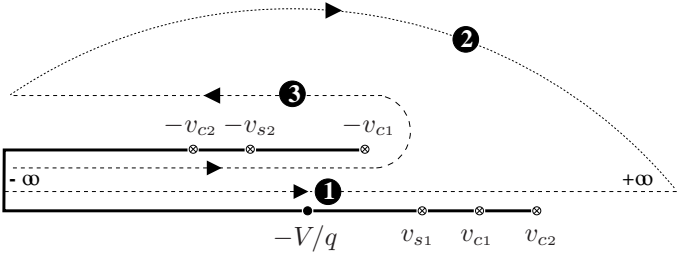}}
\caption{Integration contour and cuts for $V,q>0$. The original
  integral of $h(u)$ defined on  
$-\infty,+\infty$ (contour 1) is modified into a integral on contour
 2+3.} 
\label{fig:cuts}
\end{figure}

\begin{figure}[t]
\psfrag{q (magnetic field)}{$q$ (magnetic field)}
\psfrag{V (potential difference)}{$V$ (voltage)}
\psfrag{V/q=vs2}{\color{white} $V/q=v_{c{\rm\scriptscriptstyle 1}}$}
\psfrag{V/q=vc1}{\color{white} $\!\!\! V/q=v_{s2}$}
\psfrag{V/q=vc2}{\color{white} $V/q=v_{c{\rm\scriptscriptstyle 2}}$}
\psfrag{V/q=-vs1}{\color{white} $V/q=-v_{s1}$}
\psfrag{V/q=-vc1}{\color{white} $V/q=-v_{c{\rm\scriptscriptstyle 1}}$}
\psfrag{V/q=-vc2}{\color{white} $V/q=-v_{c{\rm\scriptscriptstyle 2}}$}
\centerline{\fig{9cm}{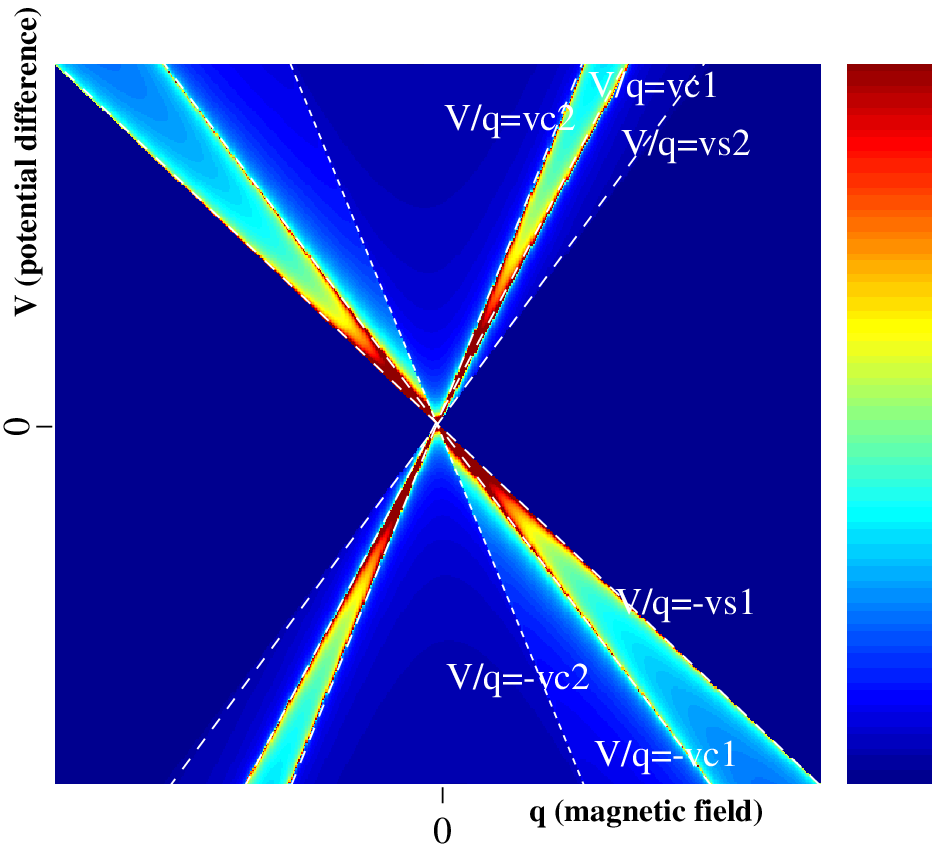}}
\psfrag{V}{$V$}
\psfrag{dJ/dV}{$dJ/dV$}
\psfrag{vs2*q}{$v_{s2}q$}
\psfrag{vc1*q}{$~v_{c{\rm\scriptscriptstyle 1}}q$}
\psfrag{vc2*q}{$v_{c{\rm\scriptscriptstyle 2}}q$}
\psfrag{-vs1*q}{$-v_{s1}q$}
\psfrag{-vc1*q}{\hspace{-.2cm}$-v_{c{\rm\scriptscriptstyle 1}}q$}
\psfrag{-vc2*q}{$-v_{c{\rm\scriptscriptstyle 2}}q$}
\centerline{\fig{8.5cm}{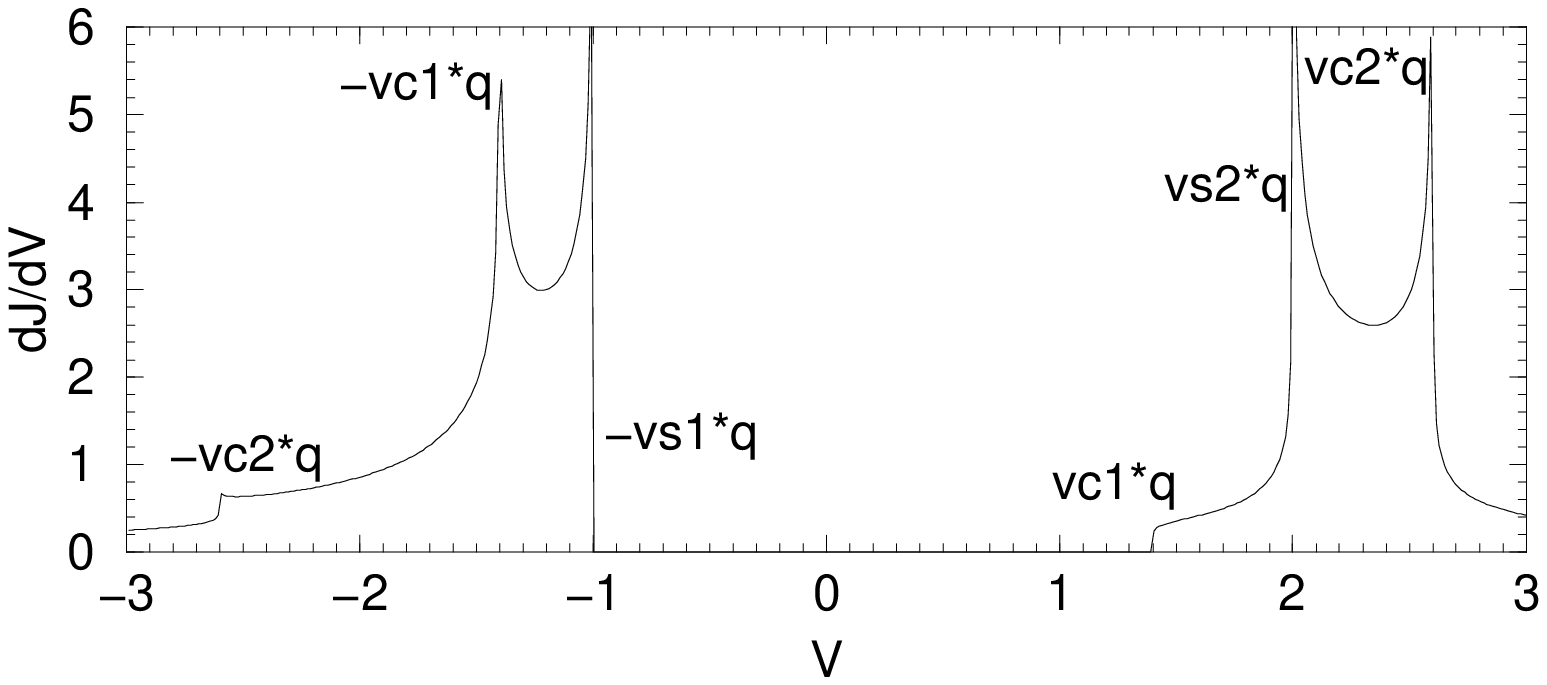}}
\caption{
  Top: Density plot of the ideal differential conductance at point A of
  Fig.~\ref{fig:nointpd}.  Bottom: Differential conductance
  $G(V,q=1)$ versus potential difference $V$.  Both plots are shown
  for values approximately appropriate to cleaved edge overgrowth:
  $\theta_{1}=0.2,\theta_2=0.32,\eta=0.18$, with velocities
  $v_{s{\rm\scriptscriptstyle 1}}=1,v_{c{\rm\scriptscriptstyle
      1}}=1.36,v_{s{\rm\scriptscriptstyle
      2}}=2,v_{c{\rm\scriptscriptstyle 2}}=2.56$ in units of $v_{\rm
    \scriptscriptstyle F1}$.}
\label{fig:plot3d}
\end{figure}

The contour of integration in (\ref{I+integral}) can then be 
deformed according to Fig.~\ref{fig:cuts}.  As 
$h(u)$ vanishes faster than $|u|^{-2}$ at infinity, we are reduced 
to the integral of $h(u)$ on the contour 3 (see Fig.~\ref{fig:cuts}). 
More explicitly in the situation where 
$v_{s{\rm\scriptscriptstyle 1}} < v_{c{\rm\scriptscriptstyle 1}} <
v_{s{\rm\scriptscriptstyle 2}} < v_{c{\rm\scriptscriptstyle 2}}$, and
for {\it e.g} $q>0$,  
the remaining integral can be 
written as the sum of real integrals:  

\begin{multline}\label{result}
J_{+}(Q,V)\propto
2 \sin(2\pi (\theta_{1}+\theta_{2})) 
\Gamma(1-2\theta_{1}-2\theta_{2})\\
\times
\sum_{\alpha=1}^{3}
\theta(V-qv_{\alpha})\sin \phi_{\alpha}
\int_{-Min(v_{\alpha+1},V/q)}^{-v_{\alpha}}
|h(u)|du , 
\end{multline}
with $v_{1}=v_{c{\rm\scriptscriptstyle
    1}},v_{2}=v_{s{\rm\scriptscriptstyle
    2}},v_{3}=v_{c{\rm\scriptscriptstyle 2}},v_{4}=+\infty$ and
$\phi_{1}=\pi(\theta_1-\eta),\phi_{2}=\pi(\theta_{1}-\eta+ \frac{1}{2}),
\phi_{3}=\pi(\theta_{1}+\theta_{2}+1/2)$.  Similar expressions for
$J_{+}$ and $J_{-}$ can be derived for any other order of the
velocities, and for the other $V\simeq 0$ points of
Fig.\ref{fig:nointpd} \cite{inprep}.  A typical result is presented in
Fig.~\ref{fig:plot3d}.

The density plot of the differential conductance (per unit length)
$G=dJ/dV$ in Fig.~\ref{fig:plot3d}\ directly exhibits evidence of
electron fractionalization.  First, non-analytic features appear along
rays (3 per quadrant) whose slope gives the two charge and spin
velocities.  Second, $dJ/dV$ is non-zero for any bias above the
threshold $|V|>v^* |q|$, with $v^* = {\rm
  Min}(v_{s2},v_{c1},v_{c2})\Theta(qV) + {\rm
  Min}(v_{s1},v_{c1},v_{c2}) \Theta(-qV)$.  Both these properties can
be understood from kinematics.  Each tunneling event corresponds to a
transfer of charge $\pm e$ and spin $\pm 1/2$ from one wire to the
other, accompanied by the addition of momentum $q$ and energy $eV$.
This creates a combination of (6 total) fractional ``chargons'' and
``spinons''.
The final state with the {\sl least} energy for a given
momentum $q$ gives {\sl all} the momentum to the {\sl slowest}
appropriate ``particle'', which determines the threshold.  Moreover,
since the final states are six-particle excitations, kinematics
allows {\sl any} energy above $v^*|q|$, explaining the non-zero weight
in $dJ/dV$ as a multi-particle continuum.  Detailed expressions for
the singularities along the various rays in Fig.~\ref{fig:plot3d}\ can
be extracted from Eq.~\ref{result}\cite{inprep}.

We now turn to a discussion of the numerous effects left out of the
above treatment.  The two most significant corrections can be treated {\sl
  exactly}.  First, the spectral density is rounded on
a scale set by temperature (the rounding can be calculated exactly
to describe detailed lineshapes\cite{inprep}).  Experimentally, for
$e^2/v_{\scriptscriptstyle F}$ of $O(1)$, LL behavior is expected to
be manifested for $eV < \epsilon_F \sim 10-20 mV$ in cleaved edge
samples.  Thus for experiments with $T$ in the Kelvin range, the
thermal rounding is minimal except at very low bias.  Second,
for $k_{{\rm\scriptscriptstyle F}a}W \gg 1$, the dominant impurity
process is elastic {\sl forward} scattering.  This is easily included
by convolving $G(V,q)$ (in $q$) calculated above with a
Lorentzian of half-width $\Delta q \equiv 1/\ell_{el}$, and has
several consequences.  The singularities at $V>0$ are rounded over an
energy width $1/\tau_{el} \sim v_{\scriptscriptstyle
  F,i}/\ell_{el}$.  Non-vanishing weight also appears outside the
kinematically allowed region.  At low bias and temperature, this
weight is itself singular and displays a pronounced {\sl low bias
  conductance dip}.  In particular, for $eV,k_{\rm \scriptscriptstyle
  B}T, v^* q \ll 1/\tau_{el}$, $G(V,q,\ell_{el}) \sim
|w|^2[\ell_{el}/(1+q^2\ell_{el}^2)] T^\beta F_\beta(eV/k_{\rm \scriptscriptstyle
  B}T)$, with $0\! <\! \beta=2\theta_1+2 \theta_2-1\!<\! 1$, where
$F_\beta(X)$ is the well-known\cite{Bockrath99}\ scaling function for {\sl
  point-contact} tunneling, satisfying $F_\beta(0)=1$ and
$F_\beta(X)\sim X^\beta$ for $X\gg 1$.

Even at $T=0$, 
effects {\sl not} included in the LL model can broaden the spectral
function.  First, consider the effects of non-forward-scattering
interactions between electrons at the edge.  In the generic situation,
$k_{\rm \scriptscriptstyle F1} \neq k_{\rm \scriptscriptstyle F2}$,
and charge/spin-density-wave coupling (e.g. ${\cal H}_{\rm CDW}=V_{\rm
  CDW} \psi^\dagger_{R1}\psi^{\vphantom\dagger}_{L1}
\psi^\dagger_{L2}\psi^{\vphantom\dagger}_{R2}$) between the two wires
is forbidden.  The dominant residual term is then the ``exchange''
interaction (we reserve the usual term ``backscattering'' for impurity
effects discussed below) within a single wire, e.g. ${\cal H}_{\rm ex,
  i} = - \lambda_i v_{\rm \scriptscriptstyle Fi} \psi_{Ri}^\dagger
\vec\sigma \psi^{\vphantom\dagger}_{Ri} \cdot \psi_{Li}^\dagger
\vec\sigma \psi^{\vphantom\dagger}_{Li}$, where $\lambda_i$
characterizes the dimensionless backscattering strength in wire $i$.
In the experimentally relevant situation, $\lambda_i \sim (k_{\rm
  \scriptscriptstyle Fi}W)^{-1} \ll 1$, so this is a weak interaction.
Formally, such exchange interactions are known to be {\sl marginally
  irrelevant} in the renormalization group sense,  
and a simple perturbative (bosonized ``phonon'' self-energy) estimate
suggests a resulting ``lifetime'' scaling approximately linearly with
energy $\omega$, 
$(1/\tau_{{\rm ex},i})^2 \sim \lambda_i^2 \omega^2/(c_1 + c_2
\lambda_i^2 \ln^2(\epsilon_{\rm \scriptscriptstyle F,i}/\omega))$,
where $c_{1/2}$ are order one constants.  
The magnitude of $1/\tau_{\rm ex,i}$, however, is small, due both to
the smallness of $\lambda_i$ and the additional logarithmic
suppression at low energies.  Other ``internal'' corrections to the
forward-scattering model, such as band curvature, are strongly
irrelevant, and a similar self-energy estimate shows that they
contribute only negligible
$1/\tau_{\rm curve} \sim O(\omega^2/\epsilon_{\rm \scriptscriptstyle
  F})$ terms to the lifetime. 
  
Electrons in the wires also interact with those in the bulk
2DEG.  
Expressing the Coulomb interaction in the basis of wire and bulk
states, one finds several distinct scattering channels.
Most interesting are {\sl charge $\pm e$ decay} processes, in which an
electron/hole in one of the wires decays into an electron/hole in the
2DEG and an electron/hole pair.  Because the 2DEG is a Fermi-liquid,
however, strong phase-space restrictions reduce the rate of such
decays at low energies: crudely, $1/\tau_{e} \sim c_e
\omega^2/\epsilon_{\rm \scriptscriptstyle F}$, where the $O(1)$
constant $c_e$ depends upon the density of the 2DEG, etc.
Interestingly, if $k_{\rm \scriptscriptstyle F,bulk}< 3 k_{\rm
  \scriptscriptstyle F,i}$, such processes are kinematically forbidden
($c_e=0$).  Similar, albeit slightly less restrictive kinematic
conditions reduce the phase space for {\sl charge $\pm 2e$ decay}, in
which Cooper pairs (hole pairs) are scattered between the wires and
2DEG.  All such processes in which charge is transferred are further
reduced by LL orthogonality catastrophe effects.  Finally, there are
forward-scattering processes in which no charge but only energy and
small momenta are transferred between the wires and 2DEG.  Because the
Coulomb potential suppresses long-wavelength charge fluctuations in
the 2DEG, we expect this also to be a weak effect, and will 
explore it in greater detail in a future publication\cite{inprep}.

In addition to the elastic forward scattering discussed above,
disorder also mixes surface and extended states, allowing conduction
at $V=T=0$.  A {\sl lower bound} on the broadening of spectral
features in the tunneling conductance is provided by the experimental
determination of the 1D-2D scattering length , $\ell_{1D-2D} \approx 6
\mu$ in Ref.~\onlinecite{Picciotto00}; hence $1/\tau_{1D-2D} > v_{\rm
  \scriptscriptstyle F,i}/\ell_{1D-2D}$.  The latter is a lower-bound
to the elastic contribution, since there is also a small-momentum
elastic component which contributes to the broadening but not
$\ell_{1D-2D}$ (or the longer elastic ``backscattering'' length
$\ell_B$ defined in Ref.~\onlinecite{Picciotto00}).  

Two additional physical effects have not been taken into account
above.  First, {\sl Zeeman coupling} leads {\sl exactly}\cite{inprep}\ 
to two superimposed copies of the LL results with $Q$ shifted left
and right by $\Delta Q^A_{\uparrow/\downarrow} = \pm \frac{g \mu_B
B}{2} (v_{\rm \scriptscriptstyle F1}^{-1} + v_{\rm 
  \scriptscriptstyle F2}^{-1})$ for point A and $\Delta
Q^B_{\uparrow/\downarrow} = \pm \frac{g \mu_B B}{2} (v_{\rm
  \scriptscriptstyle F1}^{-1} - v_{\rm \scriptscriptstyle F2}^{-1})$
for point B.  Second, {\sl direct 2DEG-wire tunneling} provides a
second, albeit weaker, conduction channel\cite{QSi}, but can be
distinguished experimentally since it contributes at low-bias for the
full range $k_{\rm \scriptscriptstyle F2}-k_{\rm \scriptscriptstyle
  F,bulk}<|Q|< k_{\rm \scriptscriptstyle F2}+k_{\rm \scriptscriptstyle
  F,bulk}$ instead of at the singular points $A,B$.

Finally, we comment on the validity of the {\sl perturbative}
treatment of tunneling.  For $q \neq 0$, the tunneling perturbation is
formally an (infinitely!) {\sl irrelevant} operator.  Thus for
sufficiently weak tunneling at non-zero $q$ the perturbative treatment
is accurate.  At $q=0$, however, tunneling is {\sl relevant}: at an
energy $\epsilon$, the effective tunneling amplitude is $w(\epsilon) =
w (\epsilon_0/\epsilon)^{3/2-\theta_1-\theta_2}$, where the cut-off
$\epsilon_0 \sim \epsilon_{\rm \scriptscriptstyle F,i}$.  Equating
this to $\epsilon_0$, we estimate the cut-off energy $eV^* \sim
\epsilon_0 (w/\epsilon_0)^{2/(3-2\theta_1-2\theta_2)}$.  For $eV,
v_{\rm \scriptscriptstyle F,i}q < eV^*$, the perturbative treatment
breaks down.  Physically, in this regime we expect {\sl coherent}
motion between the two wires.  Due to the necessary complication of
including dissipation in the leads, we therefore postpone the analysis
of this limit to a future publication\cite{inprep}.

L.B. was supported by NSF grant DMR--9985255, and the Sloan and
Packard foundations. D.C. was supported by the ITP through
NSF--DMR--9528578.  C.P. was supported by PRAXIS/BD/18554/98.

\end{multicols}

\end{document}